\documentclass[conference]{IEEEtran}

\usepackage{graphicx}
\usepackage{ulem}
\usepackage{subfigure} 
\usepackage{epsfig}
\usepackage{color}

\begin{document}
%
\title{
{\small To appear in 2013 IEEE International Symposium on Phased Array Systems \& Technology. \copyright 2013 IEEE\\~\\}
Design and Commissioning of the\\ LWA1 Radio Telescope}


\author{
\IEEEauthorblockN{S.W.\ Ellingson}
\IEEEauthorblockA{Dept.\ of Electrical \& Computer Eng.\\
Virginia Tech\\
Blacksburg VA 24061\\
ellingson@vt.edu}
\and
\IEEEauthorblockN{J.\ Craig, J.\ Dowell, G.B.\ Taylor}
\IEEEauthorblockA{Dept. of Physics \& Astronomy\\
University of New Mexico\\ 
Albuquerque NM 87131\\
\{jcraig,jdowell,gbtaylor\}@unm.edu
}
\and
\IEEEauthorblockN{J.F.\ Helmboldt}
\IEEEauthorblockA{Code 7213\\
U.S. Naval Research Laboratory\\
Washington, DC 20375, USA\\
joe.helmboldt@nrl.navy.mil
}
}


\maketitle

\begin{abstract}
LWA1 is a new large radio telescope array operating in the frequency range 10--88 MHz, located in central New Mexico.  The telescope consists of about 260 pairs of dipole-type antennas whose outputs are individually digitized and formed into beams. Simultaneously, signals from all dipoles can be recorded using one of the telescope's ``all dipoles'' modes, facilitating all-sky imaging. Notable features of the instrument include four independently-steerable beams utilizing digital “true time delay” beamforming, high intrinsic sensitivity ($\approx6$~kJy zenith system equivalent flux density), large instantaneous bandwidth (up to 78 MHz), and large field of view (about 3--10$^{\circ}$, depending on frequency and  zenith angle of pointing).  This paper summarizes the design of LWA1, its performance as determined in commissioning experiments, and results from early science observations demonstrating the capabilities of the instrument. 
\end{abstract}

%
\IEEEpeerreviewmaketitle

\section{\label{sIntro}Introduction}

\begin{figure}
\begin{center}\psfig{file=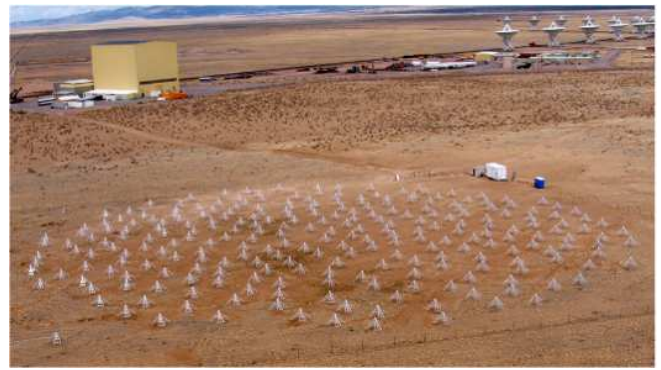,width=3.4in}\end{center} 
\caption{\label{fLWA1} LWA1.  The white structure beyond the core array is the electronics shelter.  Visible in the background is the VLA.}
\end{figure}

LWA1 (``Long Wavelength Array Station 1''; Figure~\ref{fLWA1}) is a new radio telescope operating in the frequency range 10--88 MHz, collocated with the Very Large Array (VLA; $107.63^{\circ}$~W, $34.07^{\circ}$~N) in central New Mexico.  The telescope consists of an array of about 260 pairs of dipole-type antennas whose outputs are individually digitized and formed into beams.  The principal technical characteristics of LWA1 are summarized in Table~\ref{tSpecs}.  

LWA1 is so-named because it is envisioned to be the first ``station'' of a 53-station long-baseline aperture synthesis imaging array known as the Long Wavelength Array (LWA), described in 
\cite{PIEEE_LWA}. 
Although the future of the LWA is uncertain, LWA1 was completed in Fall 2011 \cite{LWAFL} and has been operational as an ``open skies'' radio observatory under the U.S.\ National Science Foundation's ``University Radio Observatories'' program since Spring 2012.
The LWA1 Radio Observatory is currently conducting user-proposed observations supporting research on topics including pulsars, astrophysical transients, extrasolar planets, cosmology, space weather, the Sun, Jupiter, and the ionosphere.
\begin{table}
\begin{center}
\begin{tabular}{|l|l|}
\hline
Beams 	                 & 4, independently-steerable \\
\hline
Polarizations            & Dual linear \\
\hline
Tunings 	         & 2 center frequencies per beam,\\
                         & independently-selectable\\
\hline
Tuning Range 	         & 24--87~MHz \\
                         & ($>$4:1 sky-noise dominated),\\
                         & 10-88 MHz usable \\
\hline
Bandwidth                & $< \approx17$ MHz $\times$ 2 tunings $\times$ 4 beams \\
\hline
Spectral Resolution 	 & Time-domain ``voltage'' recording; also\\
                         & real-time 2048-channel spectrometer. \\
\hline
Beam FWHM 	         & $< 3.2^{\circ}\times\left[(\mbox{74 MHz})/\nu\right]^{1.5}$ \\
                         & (upper bound independent of $Z$)\\
\hline
Beam SEFD 	         & $\approx6$~kJy at $Z=0$; depends on pointing, \\ 
                         & celestial coordinates, \& frequency; \\
                         & see Table~\ref{tCygA}; also Fig.~12 of \cite{E13}  \\
Beam Sensitivity         & $\approx8$~Jy ($5\sigma$) for 1~s, 16~MHz, $Z=0$\\
                         & (inferred from SEFD)\\
\hline
All-Dipoles Modes 	 & ``TBN'': 70~kHz from every dipole,\\
                         & continuously \\
                         & ``TBW'': 78~MHz from every dipole,\\
                         & in 61~ms ``bursts'' every 5 min \\
\hline
\end{tabular}
\end{center}
{\it Notes:} $Z$ is zenith angle. $\nu$ is frequency. 1~Jy = $10^{-26}$~W~m$^{-2}$~Hz$^{-1}$. FWHM is full-width at half-maximum.  SEFD is system equivalent flux density.
~\\ 
\caption{\label{tSpecs}LWA1 Technical Characteristics.}
\end{table}

Contemporary radio telescopes which are also capable of operating in LWA1's 10--88~MHz frequency range include GEETEE (35--70~MHz), located in Gauribidanur, India \cite{GEETEE}; UTR-2 (5--40~MHz), located in the Ukraine \cite{UTR2}; VLA (74~MHz) \cite{VLA74}; and LOFAR (10--80~MHz), located in Northern Europe \cite{LOFAR13}.  

In \cite{E13} we described the high-level design of LWA1 in its Spring 2012 ``initial operational capability'' configuration (IOC), method of calibration, and measurements of beam shape and sensitivity.
In the present paper we will briefly review the current design and configuration of the instrument, which has been significantly augmented since IOC,
review beamforming performance as determined in commissioning observations,
and provide a brief summary of recent science commissioning observations.

 
\section{\label{sDesign}Design}

\begin{figure}
\begin{center}
\psfig{file=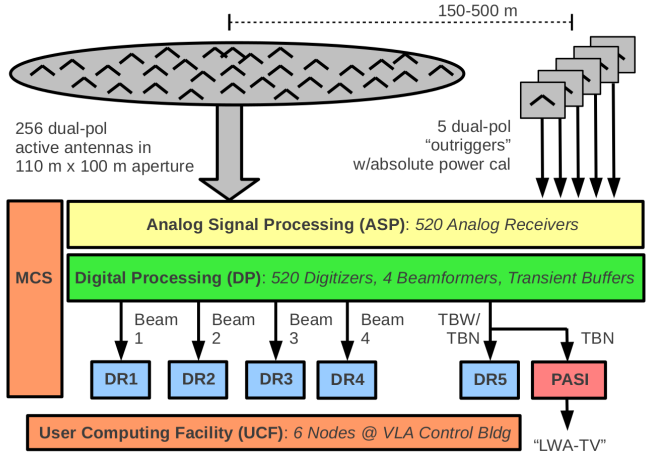,width=3.6in}
\end{center}
\caption{\label{fSysArch}
LWA1 system architecture.
}
\end{figure}

 
The LWA1 system architecture is shown in Figure~\ref{fSysArch}.  Design details are provided below, presented in order of signal flow. 

\begin{figure}
\begin{center}
\psfig{file=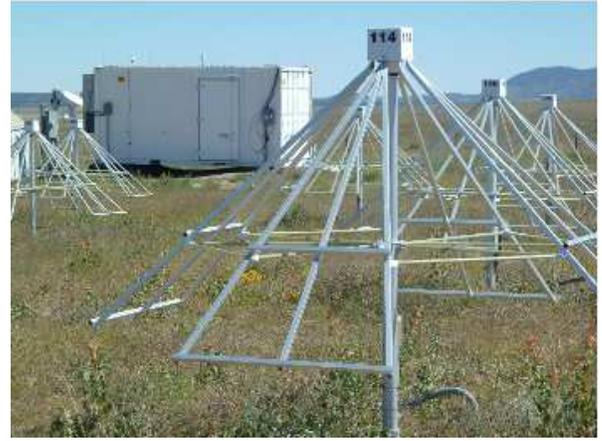,width=3.1in}
\end{center}
\caption{\label{fStand}
LWA1 antenna stands.  Front end electronics are enclosed in the white box at the feedpoint.  Signals exit through coaxial cables inside the mast. Each stand is about 1.5~m high.  The equipment shelter is visible in the background.
}
\end{figure}

{\it Antennas.} LWA1 antennas are grouped into ``stands'', each consisting of a linear-orthogonal pair of antennas, feedpoint-mounted electronics, a mast, and a ground screen as shown in Figure~\ref{fStand}.  Each antenna is a wire-grid ``bowtie'' about 3~m long, with arms bent downward at $45^{\circ}$ from the feedpoint in order to improve pattern uniformity over the sky.  The feedpoint is located 1.5~m above ground.  The ground screen is a 3~m $\times$ 3~m wire grid with spacing 10~cm $\times$ 10~cm and wire radius of about 1~mm. The primary purpose of the ground screen is to isolate the antenna from the earth ground, whose electromagnetic properties vary significantly as a function of moisture content.   
Additional details of the design and analysis of these antennas can be found in \cite{Ellingson11} and \cite{Hicks12}. 

\begin{figure}
\begin{center}
\psfig{file=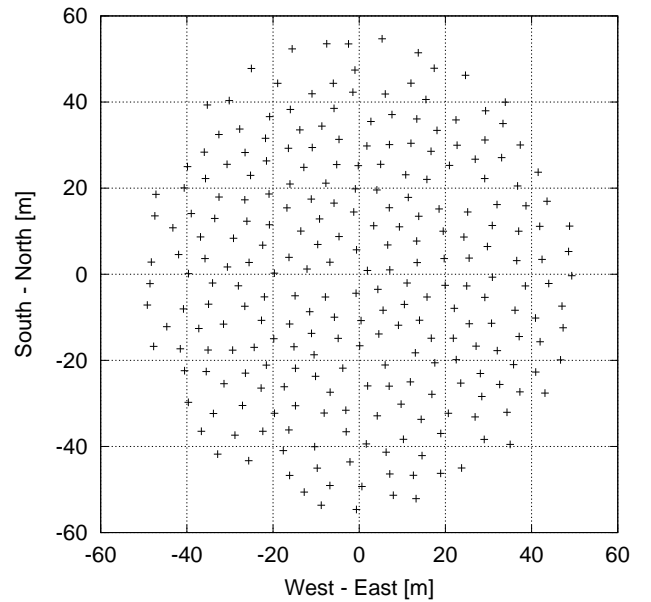,width=3.3in} 
\end{center}
\caption{\label{fAG}
Arrangement of stands in the LWA1 core array.  The minimum distance between any two masts is 5~m ($0.33\lambda$, $0.63\lambda$, and $1.23\lambda$ at 20~MHz, 38~MHz, and 74~MHz, respectively).  All dipoles are aligned North-South and East-West.  Outrigger stands are outside the limits of this view (See Figure~\ref{fAGO}.)
}
\end{figure}

{\it Array Geometry.}  LWA1 consists of 256 antenna stands (512 antennas) within a 100~m (East-West) $\times$ 110~m (North-South) elliptical footprint, plus five stands (10 antennas), known as ``outriggers'', which lie roughly 150--500~m beyond this footprint.  The arrangement of the central 256 stands forming the core array is shown in Figure~\ref{fAG}.  The physical aperture and number of stands per station were originally determined from an analysis of requirements for the LWA aperture synthesis imaging array, as detailed in \cite{PIEEE_LWA}.  However, these choices are also appropriate for the present single-station instrument.  This choice of physical aperture and number of stands results in a mean spacing between stands of about 5.4~m, which is $0.36\lambda$ and $1.44\lambda$ at 20~MHz and 80~MHz respectively.  To suppress aliasing, antennas are arranged in a pseudo-random fashion, with a minimum spacing constraint of 5~m in order to facilitate maintenance.  The elongation of the station aperture in the North-South direction improves main lobe symmetry for pointing towards lower declinations, including the Galactic center, which transits at $Z\approx63^{\circ}$ as seen from the site.
%
%

\begin{figure}
\begin{center}
\psfig{file=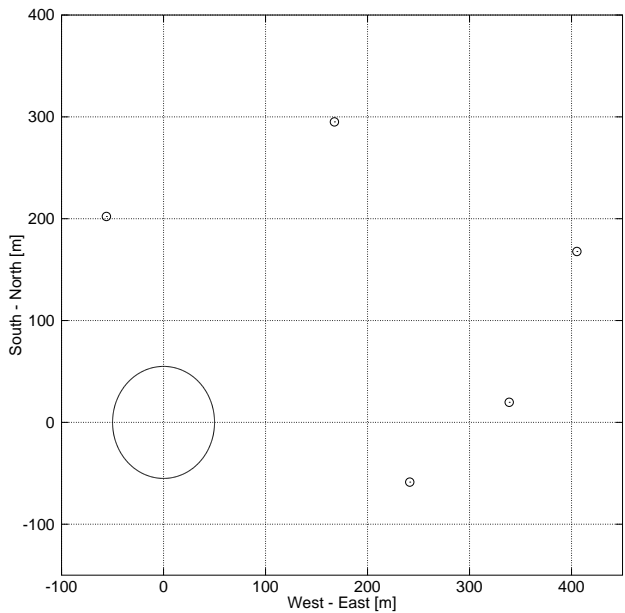,width=3.3in} 
\end{center}
\caption{\label{fAGO}
Outrigger geometry.  The ellipse in the lower left is the perimeter of the core array.
}
\end{figure}

The locations of the outrigger stands are shown in Figure~\ref{fAGO}.  These locations were selected primarily for coplanarity with the core array, and were constrained by property boundaries, forcing locations primarily to the north and east.

\begin{figure}
\begin{center}
\psfig{file=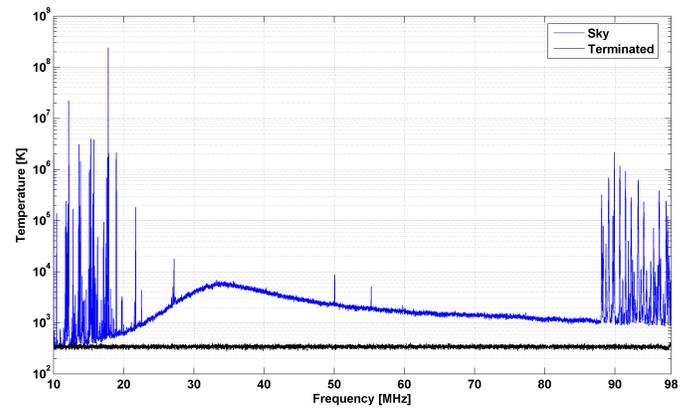,width=3.5in} 
\end{center}
\caption{\label{fFEE}
Power spectral density 
measured by an LWA1 receiver, calibrated to the antenna terminals.  
{\it Top curve}: Result expressed as system temperature 
(spikes are human-generated signals); 
{\it Bottom curve}: Same measurement made with a short circuit termination at the input, which provides an estimate of the internal contribution to the system temperature.  Spectral resolution: 6~kHz. Integration time: 10~s. Early afternoon local time.
}
\end{figure}

{\it Front End Electronics -- Core Array.}  Each dipole in the core array is terminated into a pair of commercial InGaP HBT MMIC amplifiers (Mini-Circuits GALI-74) in a differential configuration, presenting a $100\Omega$ balanced load to the antenna.  This is followed by a passive balun which produces a $50\Omega$ single-ended signal suitable for transmission over coaxial cable, plus some additional gain to overcome cable loss.  The total gain, noise temperature, and input 1~dB compression point of the resulting front end electronics units are approximately 36~dB, 300~K, and $-18$~dBm respectively, and are approximately independent of frequency over 10--88~MHz. 
%
The gain and noise temperature of the feedpoint electronics are such that they dominate the noise temperature of the complete receiver chain, which is much less than the antenna temperature 
The 1~dB compression point has been found to be satisfactory at the LWA1 site.  Although higher 1~dB compression would be better, this would be difficult to achieve without compromising system temperature.  
%
The curve labeled ``Sky'' in Figure~\ref{fFEE} shows the noise temperature measured by an LWA1 receiver in normal operation (i.e., dipole attached), after calibration to account for the known gain of the electronics following the antenna terminals.  Thus, this is an estimate of the system temperature.  Also shown is the same measurement with the dipole terminals short-circuited, which is assumed to zero the noise delivered to the front end electronics unit without significantly changing its behavior, which is consistent with the findings of laboratory experiments.  These result confirm an internal noise temperature of about 300~K, 4:1 external noise dominance over 24--87~MHz,
and negligible level of intermodulation.  
%

\begin{figure}
\begin{center}
\psfig{file=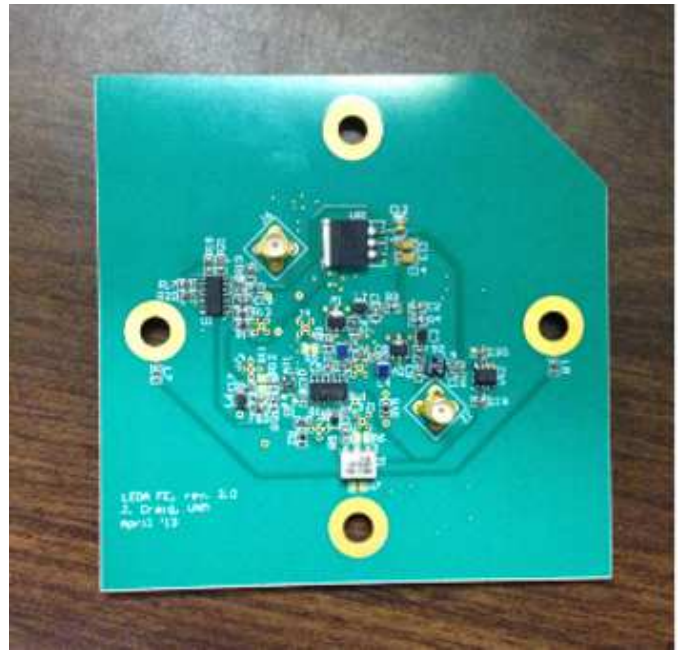,width=3.5in} 
\end{center}
\caption{\label{fFEEO}
Front end electronics board used for outrigger stands.  A heater which normally obscures the bottom left corner of the board has been removed for clarity.  One board supports both antennas in a stand.  
}
\end{figure}

{\it Front-End Electronics -- Outriggers.} The front end electronics board for outriggers is shown in Figure~\ref{fFEEO}.  The outrigger stands use front end electronics which are similar to those of the core array, but different in two respects.  First, an input impedance of $200~\Omega$ is presented to the antenna, which improves sensitivity above 45~MHz. Second, circuitry to facilitate absolute power calibration is implemented.   A three-state switching scheme is employed in which the states are (1) antenna (normal through), (2) Passive temperature-stabilized matched load (low noise temperature reference), and (3) Active noise source (high noise temperature reference).  Comparison of the measured power spectral density in each of the three states permits solution for the absolute antenna temperature.

{\it Cable System.} Connections between the front end electronics in the core array stands and the electronics shelter are by coaxial cables having lengths between 43~m and 149~m.  These cables have loss of about $-0.11$~dB/m at 88~MHz, and dispersive delays (that is, frequency-dependent delays in addition to the delay implied by the velocity factor) given by \cite{LWA170}:
\begin{equation}
\left(2.4~\mbox{ns}\right) \left( \frac{l}{100~\mbox{m}} \right) \left( \frac{\nu}{10~\mbox{MHz}} \right)^{-1/2} ~\mbox{,}
\end{equation}
where $l$ is length and $\nu$ is frequency.
Thus the signals arriving at the electronics shelter experience unequal delays, losses, and dispersion.  These can be corrected either ``in line'' in the digital processor (see below) or, for the all-dipoles modes, as a post-processing step.  In the current real-time beamforming implementation, the non-uniform dispersive delays are compensated for the center frequency of the highest-frequency tuning in a beam; thus there is some error over the bandwidth of the beam, and additional dispersion error for the lower-frequency tuning in the same beam.

\begin{figure}
\begin{center}
\psfig{file=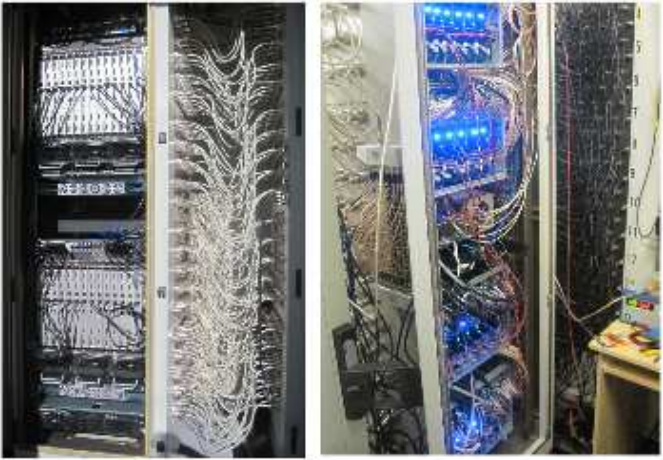,width=3.5in}
\end{center}
\caption{\label{fRacks}
Right to left: Signal entry panel, analog signal processing (ASP) rack, digital processing (DP) rack.
}
\end{figure}

{\it Equipment Shelter.}  The remaining components of the system are contained in an air-conditioned and electromagnetically-shielded shipping container, visible in Figure~\ref{fStand}.  Signals from all 520 antennas pass through a ``signal entry panel'', which connects in turn to the ``analog signal processing'' (ASP) and ``digital processing'' (DP) racks shown in Figure~\ref{fRacks} (also identified in Figure~\ref{fSysArch}).  The ASP rack contains analog receivers and circuitry for controlling front ends, whereas the DP contains digitizers and subsequent signal processing.

\begin{figure}
\begin{center}
\psfig{file=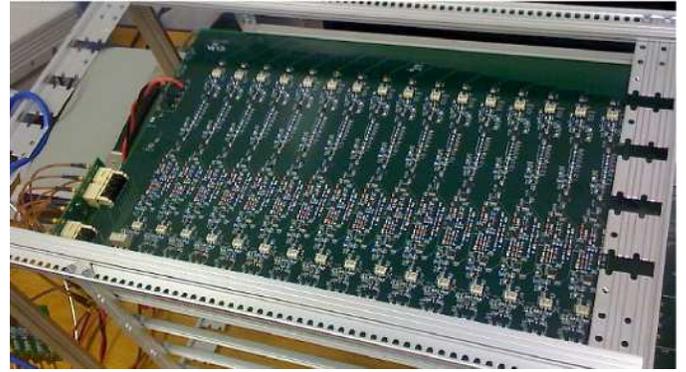,width=3.5in}
\end{center}
\caption{\label{fARX}
A view of one (of 33) analog receiver boards contained in the ASP rack.  Each board processes 16 antennas. 
}
\end{figure}

\begin{figure}
\begin{center}
\psfig{file=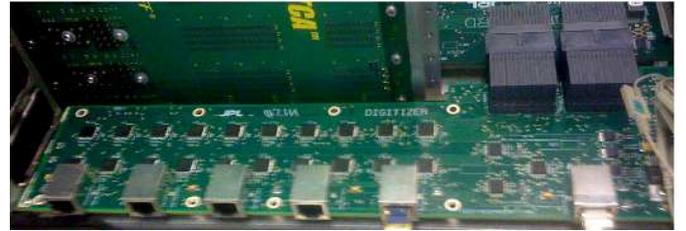,width=3.5in}
\end{center}
\caption{\label{fDIG}
One of the 26 digitizer boards contained in the DP rack.  Each board digitizes 20 antennas, and plugs into a DP1 board (see Figure~\ref{fDP1}). {\it Courtesy JPL.}
}
\end{figure}

\begin{figure}
\begin{center}
\psfig{file=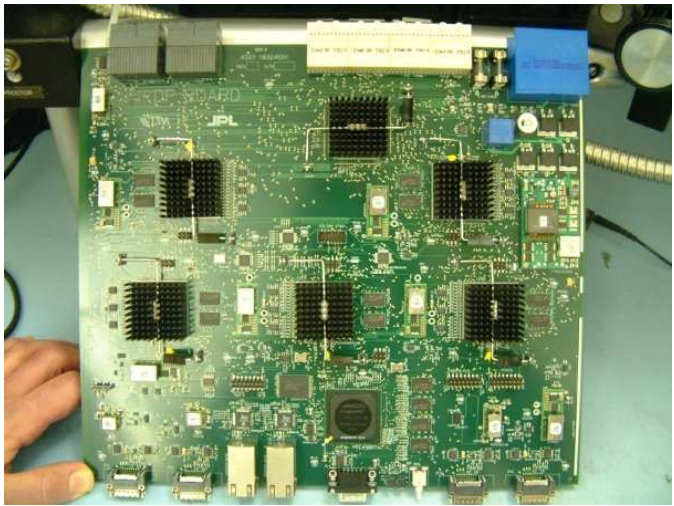,width=3.5in}
\end{center}
\caption{\label{fDP1}
One of the 26 DP1 boards contained in the DP rack.  Each DP1 board contains 5 Xilinx Virtex-5 FPGAs and implements full-bandwidth delay-and-sum beamforming and TBN/TBW for 20 dipoles. Separate ``DP2'' boards (not shown) are used for tuning of center frequencies, sample rate reduction, and transport of the beam outputs to ethernet. {\it Courtesy JPL.}
}
\end{figure}

{\it Receivers.} The signal from every antenna is processed by a direct-sampling receiver comprised of an analog section consisting of only gain and filtering (see Figure~\ref{fARX}), a 12-bit analog-to-digital converter (A/D) which samples 196 million samples per second (MSPS) (see Figure~\ref{fDIG}), and subsequent digital processing to form beams and tune within the digital passband (see Figure~\ref{fDP1}).
Digitization using fewer than 12 bits would be sufficient \cite{Memo121}, but the present design eliminates the need to implement gain control in the analog receivers and provides generous headroom to accommodate interference when it becomes anomalously large.
The choice of 196~MSPS ensures that strong radio frequency interference (RFI) from the 88--108 MHz FM broadcast band (see Figure~\ref{fFEE}) aliases onto itself, with no possibility of obscuring spectrum below 88~MHz.  
To accommodate the various uncertainties in the RFI environment, analog receivers can be electronically reconfigured between three modes:  A full-bandwidth (10--88~MHz) uniform-gain mode; a full-bandwidth dual-gain mode in which frequencies below about 35~MHz can be attenuated using a shelf filter\footnote{A ``shelf filter'' is a filter which has two adjacent passbands, with one passband (the “shelf”) having higher attenuation than the other.}; and a 28--54~MHz mode, which serves as a last line of defense should RFI above and/or below this range become persistently linearity-limiting.  In addition, the total gain in each mode can be adjusted over a 60~dB range in 2~dB steps, allowing fine adjustments to optimize the sensitivity-linearity tradeoff.  
Use of the 28--54~MHz mode has not been required to date.   
%
%

{\it Digital Beamforming.}   A detailed description of the LWA1 digital processor is provided in \cite{S+11} and is summarized here.  Beams are formed using a time-domain delay-and-sum architecture, implemented on ``DP1'' boards as shown in Figure~\ref{fDP1}. Delays are implemented in two stages: An integer-sample ``coarse'' delay is applied using a first-in first-out (FIFO) buffer operating on the A/D output samples, followed by a 28-tap finite impulse response (FIR) filter that implements an all-pass subsample delay. The filter coefficients can also be specified by the user, allowing the implementation of beams with custom shapes and nulls. The delay-processed signals are added to the signals from other antennas processed similarly to form beams. Four dual-polarization beams are constructed in this fashion, each fully-independent and capable of pointing anywhere in the sky. On separate (``DP2'') boards, each beam is subsequently converted to two independent ``tunings'' of up to 16~MHz bandwidth (4-bits ``I'' + 4-bits ``Q'' up to 19.6~MSPS) each, with each tuning having a center frequency independently-selectable from the range 10--88 MHz.  
%

Both polarizations and both tunings of a beam emerge as a single stream of user datagram protocol (UDP) packets on 10~Gb/s ethernet. 
Thus there are four ethernet output cables, with each one representing two center frequencies from a particular pointing on the sky.
The maximum data rate (ignoring protocol bits) on each ethernet cable carrying beam data is therefore 19.6 MSPS $\times$ 8~bits/sample $\times$ 2 polarizations $\times$ 2 tunings = 627.2~Mb/s.


{\it All-Sky Modes.}  Simultaneously with beamforming, LWA1 is able to coherently capture and record the output of all its A/Ds, where each A/D corresponds to one antenna. This can be done in two distinct modes. The ``transient buffer -- wideband'' (TBW) mode allows the raw ($\approx78$~MHz) 12-bit output of the A/Ds to be collected in bursts of 61 ms at a time, and $\approx5$ minutes is required to write out the captured samples. The ``transient buffer -- narrowband'' (TBN) mode, in contrast, allows a single tuning of $\approx70$~kHz bandwidth to be recorded continuously for up to 20~hours.
These two modes share the same 10 Gb/s ethernet output from the digital processor, and thus are mutually exclusive. However, the TBW/TBN output is distinct from the four beam outputs and runs simultaneously with all four beams.

{\it Data Recorders.} The limited data rate of the internet connection from the LWA1 site makes data transfer from the site impractical for observations longer than a few minutes.  Instead, each beam output and the TBW/TBN output is connected to a dedicated data recorder (DR). A DR is a computer that records the UDP packets to a ``DR storage unit'' (DRSU). 
Currently, a DRSU consists of five 3~TB drives (15 TB total) in a 1U rack-mountable chassis, configured as an eSATA disk array. 
Each DR can host 2 DRSUs, so the total data volume that can be stored on site is about 15~TB $\times$ 2 DRSUs per DR $\times$ 5 DRs = 150~TB.  At the maximum beam bandwidth, each DRSU has a capacity of about 45 hours of observation.  

{\it Real-Time Spectrometers.}  Alternatively, the DRs can be used in ``spectrometer mode'' in which they continuously compute fast Fourier transforms (FFTs) on the incoming beam data independently for each beam and tuning, and then time-average the magnitude-squared FFT output bins to the desired time resolution.  This processing is done entirely in software. 
The number of channels is selectable between 64 and 2048, independent of sample rate.  At the highest sample rate, integration times between 3~ms and 160~ms are possible depending on the number of channels selected, and these times increase proportionally with decreasing sample rate.

\begin{figure}
\begin{center}
\psfig{file=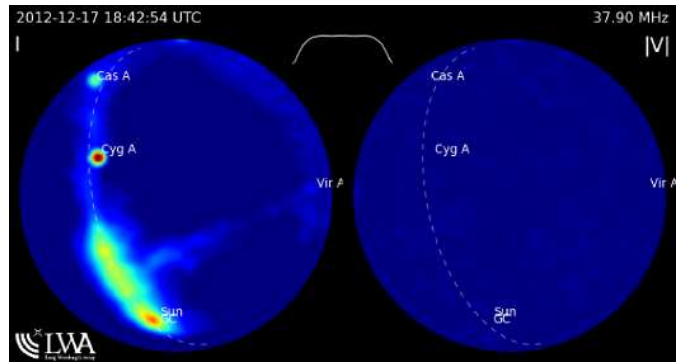,width=3.5in}
\end{center}
\caption{\label{fPASI}
An example of PASI processing of TBN output (in this case, 70~kHz bandwidth centered at 37.9~MHz), as displayed on the LWA TV website.$^{3}$
This display shows the sky in Stokes $I$ (left) and $|V|$ (right).  Note that the Galactic center, Galactic plane, and North Polar Spur are clearly visible, as are the radio galaxy Cyg~A, supernova remnant Cas~A, and the Sun.  The $|V|$ display is useful for discriminating circularly-polarized emission such as that from Jupiter.
{\it Courtesy J. Hartman (JPL).}
}
\end{figure}

{\it PASI.} The "Prototype All-Sky Imager" (PASI) is a software-defined correlator/imager currently operational at LWA1 using the TBN data stream.  
It consists of a cluster of 4 server-class computers with Nehalem multicore processors interconnected by an Infiniband switch.  
PASI images nearly the whole sky in the Stokes $I$ and $V$ parameters many times per minute, continuously and in real time, with an average duty cycle of better than 95\%.  
PASI does this by cross-correlating the dipole data streams, producing a sampling of ``visibilities'' within the station aperture.  
These visibilities are then transformed into sky images using the NRAO's Common Astronomy Software Applications (CASA) software library.\footnote{http://casa.nrao.edu/}  
Figure~\ref{fPASI} shows a sample of PASI output as it appears on the continuously-updating and publicly-accessible ``LWA TV'' website.\footnote{http://www.phys.unm.edu/$\sim$lwa/lwatv.html}

{\it User Computing Facility.}  In Fall 2012 we implemented a computer cluster in the nearby VLA Control Building known as the LWA1 User Computing Facility (UCF).  The purpose of the UCF is to provide an on-site computing facility for users of LWA1 data, eliminating the need to transport data off-site in most cases.  The UCF currently consists of six identical nodes.  Each node is a 3.20 GHz Intel Core i7-3930K CPU with 32~GB RAM and 12~TB disk storage, with Nvidia GeForce GTX6800 and Tesla C2075 GPU cards.  The cluster is connected to the DRs in the LWA1 equipment shelter via a 10~Gb/s fiber optic link, and is accessible to authorized users from any location via the internet.

\section{\label{sCal}Array Calibration and Beamforming}
 
{\it Delay Calibration.} Since LWA1 uses delay-and-sum beamforming, the primary calibration problem is to determine the set of delays which, when applied to the dipole (A/D) outputs, results in a beam with maximum directivity in the desired direction, subject to no other constraints. (In \cite{Ellingson11}, this is referred to as ``simple'' beamforming.  Other approaches are possible but have not yet been considered for implementation.)  In principle these delays can be estimated {\it a priori}, since the relevant design values (in particular, cable lengths) are known in advance.  In practice this does not work well due to errors in presumed cable lengths and the accumulation of smaller errors associated with distribution of signals within equipment racks.  Thus, delays must be determined using external stimulus signals while the instrument is in operation.  

Fortunately, it is possible to use astrophysical signals for this purpose.  Section~IV of \cite{E13} describes a method based on analysis of fringes (correlations between antennas as a function of time) in TBN data that we have found to be effective.  In that method, fringes are collected between an outrigger and each stand in the core array, requiring an observation on the order of hours to obtain instrumental phases at the TBN center frequency.  This process is repeated while varying the TBN center frequency, resulting in phase vs. frequency responses from which the desired delays can be calculated.
The primary drawback is that many days of observation are required.

For this reason we have recently developed an alternative approach using TBW, which has the advantage of having bandwidth equal to the entire tuning range.  In this approach, each TBW capture is cross-correlated to obtain visibilities, and delays are determined using self-calibration imaging over 35 to 80 MHz.  The procedure employs a simplified sky model consisting of the astrophysical sources Cyg~A, Cas~A, Tau~A, Vir~A, Her~A, Sgr~A, and the Sun, all represented as point sources.  Since the sky model neglects diffuse (Galactic) emission, self-calibration is carried out using only baselines with $uvw$ lengths greater than 14$\lambda$ at 57.5~MHz, for which diffuse emission is effectively uncorrelated.  
The self-calibration routine is then iterated until the maximum absolute deviation in delays between two consecutive iterations is less than or equal to 0.2~ns on all baselines.  
The short duration of TBW acquisitions (61~ms) is not a limitation in this approach, since sufficient signal-to-noise ratio can be achieved for the point sources considered in just a few acquisitions.

{\it Beamfoming Performance.}  The beamforming performance of LWA1 as demonstrated in commissioning experiments is documented in \cite{E13} (Section~V).  Summarizing: Beam pointing, beamwidth, and sensitivity were measured using drift scans.  A drift scan is a measurement of total power vs. time from a beam pointing in a fixed direction while a strong, unresolved astrophysical source moves through the beam.  Procedures for calculating FWHM and sensitivity are explained in \cite{E13} (Appendices~A and B, respectively).  
Because the sensitivity of LWA1 is strongly Galactic noise-limited, beam directivity, collecting area, or effective aperture are not useful metrics of sensitivity. Instead, we evaluate sensitivity in terms of system equivalent flux density (SEFD), which is defined as the flux density of an unresolved source that doubles the power spectral density at the output of the beam relative to the value in the absence of the source. 
Results from a representative drift scan measurement are shown in Table~\ref{tCygA}.   
This particular measurement is somewhat pessimistic due to the proximity of the source (Cyg A) to the Galactic plane, which causes FWHM and SEFD estimates from the drift scan method to be biased toward higher values.  However this procedure has been repeated for a variety of sources over the entire sky, with results reported in \cite{E13}; see Table~\ref{tSpecs} of the present paper for a summary.  
\begin{table}
\begin{center}
\begin{tabular}{lrl}
Frequency & FWHM & SEFD \\
\hline
85.00~MHz &  $2.8^{\circ}$ & 11.0~kJy \\
74.03~MHz &  $3.5^{\circ}$ & 16.1~kJy \\
62.90~MHz &  $4.6^{\circ}$ & 21.5~kJy \\
52.00~MHz &  $5.8^{\circ}$ & 28.8~kJy \\
37.90~MHz &  $7.5^{\circ}$ & 21.5~kJy \\
28.80~MHz &  $8.0^{\circ}$ & 18.2~kJy \\
20.50~MHz & $10.9^{\circ}$ & 47.0~kJy \\
\end{tabular}
\end{center}
\caption{
\label{tCygA}
Measured beamwidth and sensitivity for an LWA1 beam pointing at the upper culmination of Cyg A (elevation $83.3^{\circ}$, north azimuth $0^{\circ}$.). Beamwidth (FWHM) is measured in the plane of constant declination.
}
\end{table}

Presently, beam pointing is accurate to within $0.2^{\circ}$ for elevations greater than about $30^{\circ}$.
The polarization and sidelobe characteristics of the array beams are reasonable but currently not well-understood; these are the focus of current activity.



\section{\label{sCS}Commissioning and Early Science}

Aside from the characterization of LWA1 beamforming described in the previous section, a primary focus of LWA1 commissioning was demonstrating the ability to do useful science.  LWA1 ``first light'' science was reported in \cite{LWAFL}.  With declaration of IOC in April 2012, the relative roles of commissioning and science have reversed, with science operations now serving as the primary method for characterizing instrument performance.  The LWA1 Radio Observatory recently (March 2013) completed its third call for proposals (CFP).  A tally of observing projects including those from the most recent CFP is shown in Table~\ref{tScience}.  (Note this table does not include roughly 1000~h of observing exclusively for the purposes of commissioning or diagnostics.)  The remainder of this paper summarizes highlights from a few of these projects that have proven useful in commissioning the instrument. 
\begin{table}
\begin{center}
\begin{tabular}{lrr}
\hline
Science Area & Approved Projects & Hours Observed\\
\hline
Hot Jupiters &  2 & 1251 \\
Pulsars      &  7 &  569 \\
Transients   &  6 &  561 \\
Sun          &  3 &  434 \\
Ionosphere   & 10 &  262 \\
Jupiter      &  3 &  214 \\
Cosmology    &  2 &   16 \\
Others       &  9 &  265 \\
\hline
TOTAL        & 42 & 3572 \\
\end{tabular}
\end{center}
\caption{\label{tScience}LWA1 Science Operations Status as of June 2013.}
\end{table}

\begin{figure}
\begin{center}
\psfig{file=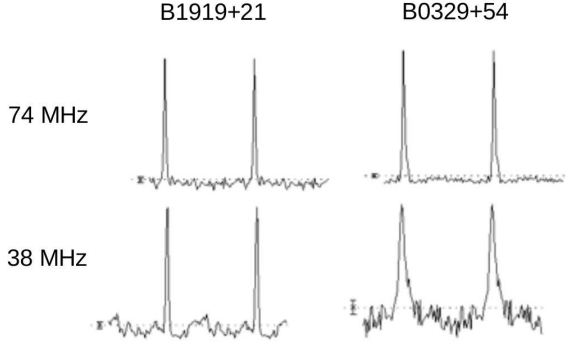,width=3in} 
\end{center}
\caption{
\label{fPulsars} 
Integrated pulse profiles for two bright pulsars.
}
\end{figure}

\begin{figure}
\begin{center}
\psfig{file=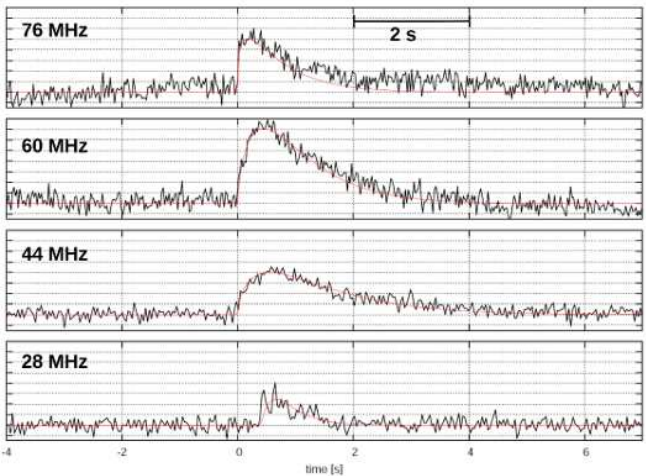,width=3.4in} 
\end{center}
\caption{
\label{fCGP} 
A giant pulse from the Crab Nebula pulsar (B0531+21) observed simultaneously in passbands of 16~MHz bandwidth 
distributed over 20--84~MHz.
Dispersion and dispersive delay between passbands has been removed, leaving only the pulse broadening associated with scattering by the interstellar medium. 
The vertical scales are unmodified from the raw data, and thus include instrumental variations in gain and sensitivity between passbands. 
}
\end{figure}

\begin{figure}
\begin{center}
\psfig{file=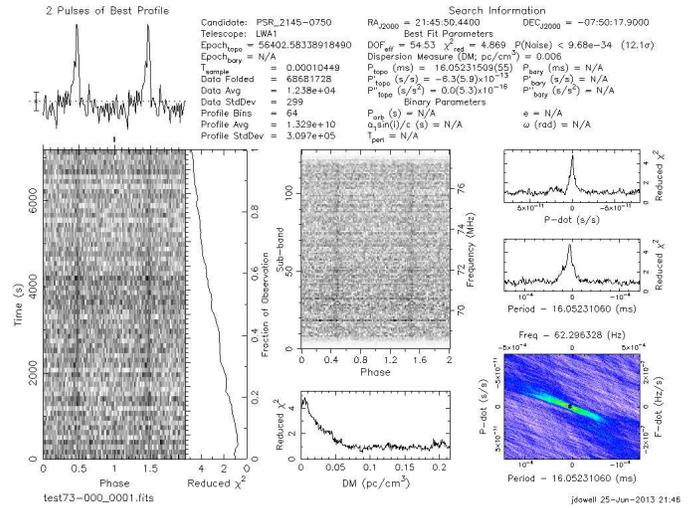,width=3.5in} 
\end{center}
\caption{
\label{fmspulsar} 
Diagnostic plot indicating detection of the millisecond pulsar J2145-0750 by LWA1 in an 2~h $\times$ 8~MHz bandwidth observation at 73~MHz. 
}
\end{figure}

LWA1 is a particularly effective instrument for low-frequency study of pulsars.  A total of 24 pulsars have been detected at LWA1 to date, as shown in Table~\ref{tPulsars}.  This list includes 22 ``normal'' pulsars detected by incoherent dedispersion and folding.  Multifrequency pulse profiles for two well-known bright pulsars are shown in Figure~\ref{fPulsars}.  The so-called ``giant pulses'' of the Crab nebula pulsar (B0531+21) are detectable throughout the frequency range of LWA1, as shown in Figure~\ref{fCGP} \cite{CGP13}. 
The millisecond pulsar J2145-0750 has been detected using coherent dedispersion at frequencies from 37~MHz to 85~MHz, demonstrating that LWA1 is able to productively study this important class of pulsars \cite{JD13}.  
Figure~\ref{fmspulsar} shows the detection of this pulsar at 73~MHz.
\begin{table}
\begin{center}
\begin{tabular}{lll}
B0031-07 & B0809+74 & B1642-03 \\
B0320+39 & B0818-13 & B1839+56 \\
B0329+54 & B0823+26 & B1919+21\\
B0450+55 & B0834+06 & B1929+10  \\
B0525+21 & B0919+06 & B2110+27 \\
B0531+21 (GPs) & B0943+10 & J2145-0750 (ms)\\
         & B0950+08 & B2217+47\\
         & B1133+16 & \\
         & B1237+25 & \\
         & B1508+55 & \\
         & B1540-06 & \\
\end{tabular}
\end{center}
\caption{\label{tPulsars}Pulsars detected by LWA1 as of June 2013.}
\end{table}

\begin{figure}
\begin{center}
\psfig{file=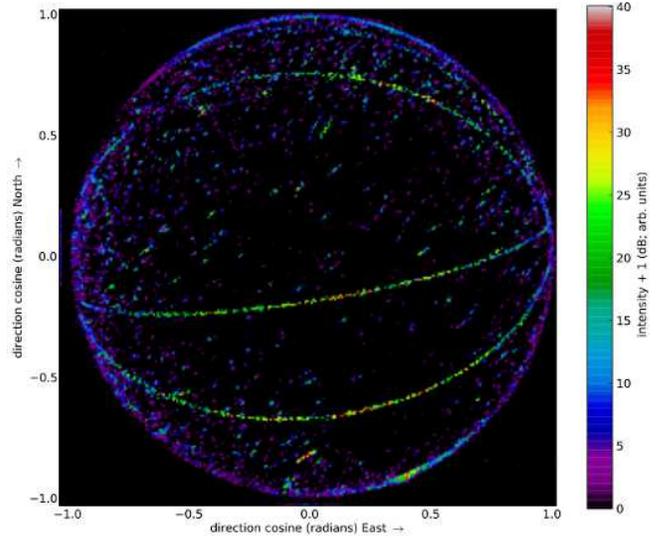,width=3.4in} 
\end{center}
\caption{
\label{fJoeH} 
An all-sky 3-hour ``time exposure'' using TBN ($\approx70$~kHz bandwidth) at 55.25~MHz, which is the center frequency of the NTSC (analog television) channel 2 video carrier.  
The continuous horizon-to-horizon tracks are reflections from aircraft passing overhead.  The bright spot on the horizon in the lower right is the groundwave from television station XEMP in Juarez, Mexico.  Many of the remaining features are reflections from meteor ionization trails.
}
\end{figure}

Non-astronomical applications of the LWA1 include study of the Earth's ionosphere, meteor ionization trails, and propagation by reflection from aircraft.  A vivid example is shown in Fig.~\ref{fJoeH}.  Recent results on ionospheric characterization and meteor trails are documented in \cite{JH-RS-OTHR} and \cite{JH-RS-MT}, respectively.

LWA1's combination of high sensitivity, large bandwidth, and high time resolution has also resulted in significant progress in the study of Jupiter and the Sun; documentation of this work is in preparation.  Preliminary examples of work in these and other areas can be found on the website http://lwa1.info.

\section{\label{sCS}Additional Information \& Proposal Solicitations}

Additional information on LWA1 engineering and science operations can be found at the project web sites http://lwa.unm.edu and http://lwa1.info, and in the LWA Memo Series, http://www.ece.vt.edu/swe/lwa/.  The project web sites also include information on proposal submission, which is possible through periodic calls for proposal, or in some cases by contacting observatory management directly.  Interested users are encouraged to contact the authors for more information.


\section*{Acknowledgments}

The authors acknowledge contributions to the design and commissioning of LWA1 made by personnel from the following organizations:  Burns Industries, NASA Jet Propulsion Laboratory (JPL), U.S. Naval Research Laboratory (NRL), University of New Mexico (UNM), University of Texas at Brownsville, and Virginia Tech (VT).
Construction of LWA1 was supported by the Office of Naval Research under Contract N00014-07-C-0147. 
Support for operations and continuing development of LWA1 is provided by the National Science Foundation under grants AST-1139963 and AST-1139974 of the University Radio Observatories program. 
Additional support for LWA1 technical development is provided by the National Science Foundation under grant AST-1106054.
The LWA1 User Computing Facility is an internally-funded joint project of JPL, UNM, and VT.
Basic research in radio astronomy at NRL is supported by 6.1 base funding.
The authors acknowledge the support of the National Radio Astronomy Observatory.



%

\end{document}